\input{aipcheck}

\documentclass[
    ,final            
  ]
  {aipproc}

\layoutstyle{8x11single}

\def\beq{\begin{equation}}
\def\eeq{\end{equation}}
\def\beqa{\begin{eqnarray}} 
\def\eeqa{\end{eqnarray}}

\begin{document}

\title{Stability Conditions For a Noncommutative Scalar Field Coupled to Gravity via the Positive Energy Theorem}


\author{Orfeu Bertolami\footnote{Also at Instituto de Plasmas e Fus\~{a}o Nuclear, Instituto Superior T\'ecnico,
Lisboa. E-mail address: orfeu@cosmos.ist.utl.pt}}{
  address={Instituto Superior T\'ecnico, Departamento de F\'\i sica, \\
Av. Rovisco Pais, 1049-001 Lisboa, Portugal }
}

\author{Carlos A. D. Zarro\footnote{Also at Instituto de Plasmas e Fus\~{a}o Nuclear, Instituto Superior T\'ecnico,
Lisboa. E-mail address: carlos.zarro@ist.utl.pt}}{
  address={Instituto Superior T\'ecnico, Departamento de F\'\i sica, \\
Av. Rovisco Pais, 1049-001 Lisboa, Portugal }
}

\begin{abstract}
The stability requirements for a noncommutative scalar field coupled to gravity is investigated through the positive energy theorem. It is shown that for a noncommutative scalar with a polynomial potential, the stability conditions are similar to the ones for the commutative case. This result remains valid even whether the space-time has horizons.  
\end{abstract}

\maketitle

\section{Introduction}
Noncommutativity is believed to be an important ingredient in the description of space-time at quantum gravity scale, presumably at order of the Planck length \cite{Connes:1996gi}. In String theory, for instance, it is shown that noncommutativity arises under certain  conditions \cite{Seiberg:1999vs}. Besides this relevance, one admits that noncommutativity is interesting on its own grounds, and can be implemented at quantum mechanics level at configuration or at full phase-space level \cite{Snyder:1947,Gamboa:2001fg,Zhang:2004yu,Bertolami:2005jw,Acatrinei:2003id,Bertolami:2005ud,Bastos:2006kj,Bastos:2006ps}, and in quantum field theories  \cite{Szabo:2001kg, Douglas:2001ba}. Other issues associated to noncommutative geometry involve the breaking of Lorentz symmetry \cite{Bertolami:1999da,Bertolami:2003nm,Carroll:2001ws}, noncommutative fields in Friedmann-Lema\^{i}tre-Robertson-Walker spaces \cite{Lizzi:2002ib,Bertolami:2002eq}, astrophysics \cite{Bertolami:2009}, black-holes \cite{Bastos:2009ae} and noncommutative quantum cosmology \cite{GarciaCompean:2001wy,Barbosa:2004kp,Bastos:2007bg}.

In this contribution, one tackles the problem of stability of a noncommutative scalar field in a general curved space-time \cite{Bertolami:2008zv}. This problem  is examined using the positive energy theorem, which states that the gravitational energy cannot be negative if matter fields satisfy the dominant energy condition \cite{Witten:1981mf,ChoquetBruhat:1985xy,Nester:1982tr}. This proves the stability of Minkowski space-time at classical and semi-classical levels. One can extend this theorem to include scalar and vector fields and this set-up can be used to show the stability of supergravity theories \cite{Gibbons:1983aq}. This method can be generalized to fields that do not admit a supersymmetric extension  \cite{Boucher:1984yx} and  this in turn, can be used to obtain the stability condition for scalar fields that are non-minimally coupled to gravity \cite{Bertolami:1987wj}. The extension of this theorem to include black holes was proposed in Ref. \cite{Gibbons:1982jg}.

The approach discussed here consists in obtaining the stability conditions for a noncommutative scalar fields defined in a curved space-time using the method of Refs. \cite{Boucher:1984yx,Bertolami:1987wj}. Non-commutativity is implemented via the Moyal product adapted to curved spaces \cite{Lizzi:2002ib}, and  a condition to ensure associativity at a given order in the noncommutative parameter $\theta$, for a noncommutative  scalar field with a polynomial potential \cite{Bertolami:2002eq}.

This work embodies the results of Ref. \cite{Bertolami:2008zv} and is organized as follows: first, one presents the model for a noncommutative scalar field and how it can be coupled to gravity. The noncommutative version of the generalized positive theorem is then obtained and the stability conditions are found for spaces with and without horizons. Finally, one presents a set of conclusions and discusses its implications.


\section{The model}\label{sec:model}
The set-up to study the stability conditions involve the following assumptions:

{\bf Assumption 1}: Noncommutativity is implemented via a covariant version of the Moyal product \cite{Lizzi:2002ib, Bertolami:2002eq, BarcelosNeto:2002bh, Harikumar:2006xf}

\beq \label{eq:covmoyal}
f \star g=\sum_{n=0}^{\infty}\frac{(i/2)^{n}}{n!}\theta^{\alpha_{1}\beta_{1}}\cdots\theta^{\alpha_{n}\beta_{n}}\left(\nabla_{\alpha_{1}}\cdots\nabla_{\alpha_{n}}f\right)\left(\nabla_{\beta_{1}}\cdots\nabla_{\beta_{n}}g\right),
\eeq

\noindent where f and g are in general tensor fields, whose indices are omitted for simplicity. The operator $\nabla_{\mu}$ denotes covariant derivative. This deformed product is not associative since the covariant derivatives do not commute. Furthermore, $\theta^{\mu\nu}$, the noncommutative parameter, is regarded as a tensor and it is given by 

\beq
[x^{\mu},x^{\nu}]=i\theta^{\mu\nu}.
\eeq

{\bf Assumption 2}: $\theta^{\mu\nu}$ is covariantly constant

\beq\label{eq:covtheta}
\nabla_{\alpha}\theta^{\mu\nu} = 0,
\eeq

\noindent this condition is a generalization of the constant $\theta^{\mu\nu}$, of the Minkowski space-time \cite{Lizzi:2002ib, Harikumar:2006xf}.

{\bf Assumption 3}: The noncommutative polynomial scalar potential is given by \cite{Bertolami:2002eq}:

\beq \label{eq:ncpot}
\tilde{V}(\Phi)=\sum_{n=0}^{\infty}\frac{\lambda_{n}}{n!}\overbrace{\Phi\star\cdots\star\Phi}^{n\;\;times},
\eeq 

\noindent where the tilde denotes a noncommutative function. Eq. (\ref{eq:ncpot}) is obtained from the polynomial scalar potential $V(\Phi)=\sum_{n=0}^{\infty}\frac{\lambda_{n}}{n!}\Phi^{n}$ by substituting the usual point-wise product  between functions by the Moyal one.

{\bf Assumption 4}: One can define an associativity condition  

\beq \label{eq:asscond}
\theta^{\mu\nu}\nabla_{\nu}\Phi=0.
\eeq

\noindent Although the covariant Moyal product is in general nonassociative, this condition keeps Eq. (\ref{eq:ncpot}) associative up to second order in the noncommutative parameter \cite{Bertolami:2002eq}. Expanding Eq. (\ref{eq:ncpot}) up to the second order in $\theta$, one finds \cite{Bertolami:2002eq}:

\beq \label{eq:ncpotexp}
\tilde{V}(\Phi)= V(\Phi) +\frac{1}{2}\frac{d^{2}V(\Phi)}{d\Phi^{2}}\left(-\frac{1}{8}\theta^{\alpha_{1}\beta_{1}}\theta^{\alpha_{2}\beta_{2}}\nabla_{\alpha_{1}}\nabla_{\alpha_{2}}\Phi \nabla_{\beta_{1}}\nabla_{\beta_{2}}\Phi \right).
\eeq

\noindent Notice that Eq. (\ref{eq:asscond}) has two classes of solutions. Choosing that $\det\theta^{\mu\nu}= 0$ does not trivialize the problem; if however, one chooses $\det\theta^{\mu\nu}\neq0$, it implies that $\nabla_{\nu}\Phi=0$, which is a too stringent condition.

{\bf Assumption 5}: The gravity sector of the model is unaffected by noncommutativity, so that the usual Einstein's equations remains unchanged,

\beq \label{eq:eeq}
G_{\mu\nu} = \kappa \tilde{T}_{\mu\nu},
\eeq

\noindent where $\kappa =8\pi G$ and $\tilde{T}_{\mu\nu}$ is the noncommutative energy-momentum tensor that comprises a sum of two components: one related to the scalar field and the other related to matter fields, $\tilde{T}_{\mu\nu}=\tilde{T}^{\Phi}_{\mu\nu}+\tilde{T}^{\mbox{M}}_{\mu\nu}$.

{\bf Assumption 6}: The matter fields satisfy the dominant energy condition\footnote{Physically this condition states that local energy density must be  positive, that is for any time-like vector $W^{\mu}$, $T_{\mu\nu}W^{\mu}W^{\nu}\geq 0$, and $T_{\mu\nu}W^{\mu}$ is not a space-like vector  \cite{Hawking:1973uf}.}. The noncommutative action then reads

\beq
\tilde{S}=\int d^{4}x\sqrt{-g}\left[\frac{1}{2}g^{\mu\nu}\nabla_{\mu}\Phi\star\nabla_{\nu}\Phi - \tilde{V}(\Phi) + \tilde{\mathcal{L}}_{\mbox{M}} \right],
\eeq

\noindent and the noncommutative energy-momentum tensors are given by

\beq
\tilde{T}^{\mbox{M}}_{\mu\nu} = \frac{2}{\sqrt{-g}}\frac{\delta\left(\sqrt{-g}\tilde{\mathcal{L}}_{\mbox{M}}\right)}{\delta g^{\mu\nu}}, \;\;\tilde{T}^{\Phi}_{\mu\nu} = \frac{1}{2}\left(\nabla_{\mu}\Phi\star\nabla_{\nu}\Phi + \nabla_{\nu}\Phi\star\nabla_{\mu}\Phi\right) -\frac{1}{2} g_{\mu\nu}\nabla_{\rho}\Phi\star\nabla^{\rho}\Phi + g_{\mu\nu}\tilde{V}(\Phi). \label{eq:ncemtsf}
\eeq

{\bf Assumption 7}: The product between spinor fields and gamma matrices is the usual one and the spinor fields commute with the noncommutative scalar field.  The 4-momentum of the gravitational field can be associated to a four-momentum vector $p_{\mu}$ in a asymptotically flat space by the following relationship \cite{Nester:1982tr}:

\beq\label{eq:wi}
16\pi G p_{\mu}V^{\mu} = \frac{1}{2}\oint_{S=\partial\Sigma} E^{\sigma\alpha}dS_{\sigma\alpha} = \int_{\Sigma}\nabla_{\alpha}E^{\sigma\alpha}d\Sigma_{\sigma},
\eeq

\noindent where $V^{\mu}=\overline{\epsilon_{0}}\gamma^{\mu}\epsilon_{0}$, $\epsilon_{0}$ represents a constant Dirac spinor, $\Sigma$ is an arbitrary three-dimensional hypersurface and $S$ its boundary  $\partial\Sigma$ at infinity. The two-form $E^{\sigma\alpha}$ is defined as\footnote{Our conventions are the following: the metric signature is $(+,-,-,-)$, $\overline{\epsilon}=\epsilon^{\dagger}\gamma^{0}$, $\{\gamma^{\mu}$, $\gamma^{\nu}\}= 2g^{\mu\nu}$, $\sigma^{\mu\nu}=\frac{1}{4}[\gamma^{\mu},\gamma^{\nu}]$, $\epsilon_{0123}=+1$, $\nabla_{\alpha}\epsilon=\partial_{\alpha}\epsilon - \frac{1}{2}\omega^{\mu\nu}_{\alpha}\sigma_{\mu\nu}\epsilon$, $\Gamma^{\sigma\alpha\beta}=\gamma^{[\sigma}
\gamma^{\alpha}\gamma^{\beta]}$, $\Gamma^{\sigma\alpha}=\gamma^{[\sigma}
\gamma^{\alpha]}$.}  

\beq \label{eq:ntf}
E^{\sigma\alpha}=2\left(\overline{\epsilon}\Gamma^{\sigma \alpha\beta}\nabla_{\beta}\epsilon - \overline{\nabla_{\beta}\epsilon}\Gamma^{\sigma \alpha\beta}\epsilon\right),
\eeq

\noindent where $\epsilon$ is a Dirac spinor which behaves as $\epsilon \rightarrow \epsilon_{0} + \mathcal{O}\left(\frac{1}{r}\right)$ at infinity. Hence, this assumption is justified as the total energy-momentum can be written with the use of spinor fields and Assumption 5.


\section{Generalized positive energy theorem}\label{sec:gpet}

As already mentioned, the positive energy theorem can be generalized in order to include fields that do not admit a supersymmetric extension. Since this extension is based on the generalized positive energy theorem, its derivation for supersymmetric theories is presented.  One defines a generalized two-form $\hat{E}^{\sigma\alpha}$ (Eq. (\ref{eq:ntf}))

\beq \label{eq:cntf}
\hat{E}^{\sigma\alpha}=2\left(\overline{\epsilon^{i}}\Gamma^{\sigma \alpha\beta}\hat{\nabla}_{\beta}\epsilon^{i} - \overline{\hat{\nabla}_{\beta}\epsilon^{i}}\Gamma^{\sigma \alpha\beta}\epsilon^{i}\right),
\eeq

\noindent where $\hat{\nabla}_{\mu}$ is the supercovariant derivative related to the change of the gravitino field under a supersymmetric transformation and $i=1,\ldots,N$ is the number of supersymmetries. One can show that Eq. (\ref{eq:wi}) is modified to

\beq \label{eq:gwi}
16\pi G p_{\mu} \overline{\epsilon_{0}^{i}}\gamma^{\mu}\epsilon_{0}^{i} = \int_{\Sigma}\left[16\pi G T^{\mbox{M}\sigma}_{\;\;\;\alpha}\overline{\epsilon^{i}}\gamma^{\alpha}\epsilon^{i}  + 4\overline{\hat{\nabla}_{\alpha}\epsilon^{i}}\Gamma^{\sigma \alpha\beta}\hat{\nabla}_{\beta}\epsilon^{i} + \overline{\delta\chi^{a}}\gamma^{\sigma}\delta\chi^{a}   \right]d\Sigma_{\sigma},
\eeq

\noindent where $\delta\chi^{a}$ represents the change of spin-$\frac{1}{2}$ fields under a supersymmetric transformation. 
The first term in the integrand of Eq. (\ref{eq:gwi}) is positive since $T^{\mbox{M}\sigma}_{\;\;\;\alpha}$ satisfies the dominant energy condition and $\overline{\epsilon_{0}^{i}}\gamma^{\alpha}\epsilon_{0}^{i}$ is non-space-like. Considering ``0'' as the time direction orthogonal to $\Sigma$, hence the last two terms of the R.H.S. of Eq. (\ref{eq:gwi}) can be written as\footnote{Latin indices span over $1,2,3$.}

\begin{equation}
4\overline{\hat{\nabla}_{m}\epsilon^{i}}(\gamma^{0}\sigma^{mn}+\sigma^{mn}\gamma^{0})\hat{\nabla}_{n}\epsilon^{i} + \left(\delta\chi^{a}\right)^{\dagger}\delta\chi^{a} = -4g^{mn}\left(\hat{\nabla}_{m}\epsilon^{i}\right)^{\dagger}\hat{\nabla}_{n}\epsilon^{i} + 4\left(\hat{\nabla}_{m}\epsilon^{i}\right)^{\dagger}\gamma^{m}\gamma^{n}\hat{\nabla}_{n}\epsilon^{i} + \left(\delta\chi^{a}\right)^{\dagger}\delta\chi^{a}.\label{eq:15}
\end{equation}

If one chooses the Witten condition \cite{Gibbons:1983aq}

\beq\label{eq:witten}
\gamma^{n}\hat{\nabla}_{n}\epsilon^{i} = 0,
\eeq

\noindent Eq. (\ref{eq:15}) is positive, and therefore the generalized positive energy theorem is proved. This method always works for supersymmetric theories since the values of $\hat{\nabla}_{n}\epsilon^{i}$ and $\delta\chi^{a}$ are set by supersymmetry \cite{Gibbons:1983aq,Boucher:1984yx}. For non-supersymmetric scalar fields, the proof of the generalized theorem is similar \cite{Boucher:1984yx}, but for the case under study one has to introduce noncommutative scalar functions that have to be found in order to ensure the positive energy theorem. One defines, generalizing the result of Ref. \cite{Bertolami:1987wj},

\begin{equation}
\hat{\nabla}_{\mu}\epsilon^{i} = \nabla_{\mu}\epsilon^{i} + \frac{i}{2}\kappa \gamma_{\mu}\tilde{f}^{ij}(\Phi)\epsilon^{j}, \;\;\delta\chi^{a} = i\gamma^{\mu}\nabla_{\mu}\Phi\star\tilde{f}^{ai}_{2}(\Phi)\epsilon^{i}+\tilde{f}^{ai}_{3}(\Phi)\epsilon^{i}, \label{eq:defchi}
\end{equation}

\noindent where $\tilde{f}^{ij}(\Phi)$, $\tilde{f}^{ai}_{2}(\Phi)$ and $\tilde{f}^{ai}_{3}(\Phi)$ are noncommutative real scalar functions to be determined. Using the spinor identity $[\nabla_{\mu},\nabla_{\nu}]\epsilon=\frac{1}{2}R^{\alpha\beta}_{\;\;\mu\nu}\sigma_{\alpha\beta}\epsilon$ and Eqs. (\ref{eq:eeq}) and (\ref{eq:ncemtsf}), one gets

\begin{eqnarray}\label{eq:ncdiv}
\nabla_{\alpha}\hat{E}^{\sigma\alpha} &=& 2\kappa\tilde{T}^{\mbox{M}\sigma}_{\;\;\;\alpha}\overline{\epsilon^{i}}\gamma^{\alpha}\epsilon^{i}  + 4\overline{\hat{\nabla}_{\alpha}\epsilon^{i}}\star\Gamma^{\sigma \alpha\beta}\hat{\nabla}_{\beta}\epsilon^{i} + \overline{\delta\chi^{a}}\star\gamma^{\sigma}\delta\chi^{a}  \nonumber \\
&+& \left(\tilde{f}^{ai}_{2}(\Phi)\star\nabla_{\alpha}\Phi\right)\star\left(\nabla_{\beta}\Phi\star\tilde{f}^{aj}_{2}(\Phi)\right)\overline{\epsilon}^{i}\Gamma^{\sigma \alpha\beta}\epsilon^{j} \nonumber  \\
&+&\left\{2\kappa\delta^{ij}\left[\frac{\nabla^{\sigma}\Phi\star\nabla_{\alpha}\Phi + \nabla_{\alpha}\Phi\star\nabla^{\sigma}\Phi}{2} - \frac{\delta^{\sigma}_{\;\;\alpha}\nabla^{\rho}\Phi\star\nabla_{\rho}\Phi}{2} \right] \right. \nonumber \\
&-&\left[\left(\tilde{f}^{ai}_{2}(\Phi)\star\nabla_{\alpha}\Phi\right)\star\left(\nabla^{\sigma}\Phi\star\tilde{f}^{aj}_{2}(\Phi)\right) + \left(\tilde{f}^{ai}_{2}(\Phi)\star\nabla^{\sigma}\Phi\right)\star\left(\nabla_{\alpha}\Phi\star\tilde{f}^{aj}_{2}(\Phi)\right) \right. \nonumber \\
&-&\left. \left. \delta^{\sigma}_{\;\;\alpha}\left(\tilde{f}^{ai}_{2}(\Phi)\star\nabla^{\rho}\Phi\right)\star\left(\nabla_{\rho}\Phi\star\tilde{f}^{aj}_{2}(\Phi)\right)\right]\right\}\overline{\epsilon}^{i}\gamma^{\alpha}\epsilon^{j} + i\left[4\kappa\nabla_{\alpha}\tilde{f}^{ij}(\Phi)  \right. \nonumber \\
&-& \left. \left(\tilde{f}^{ai}_{2}(\Phi)\star\nabla_{\alpha}\Phi\right)\star\tilde{f}^{aj}_{3}(\Phi) - \tilde{f}^{ai}_{3}(\Phi)\star\left(\nabla_{\alpha}\Phi\star\tilde{f}^{aj}_{2}(\Phi)\right)\right]\overline{\epsilon}^{i}\Gamma^{\sigma\alpha}\epsilon^{j} \nonumber \\
&+&\left[-\tilde{f}^{ai}_{3}(\Phi)\star\tilde{f}^{aj}_{3}(\Phi)+2\kappa\delta^{ij}\tilde{V}(\Phi)+ 6\kappa^{2}\tilde{f}^{il}(\Phi)\star\tilde{f}^{lj}(\Phi)\right]\overline{\epsilon}^{i}\gamma^{\sigma}\epsilon^{j}  \nonumber 	\\
&+&i\left[\left(\tilde{f}^{ai}_{2}(\Phi)\star\nabla^{\sigma}\Phi\right)\star\tilde{f}^{aj}_{3}(\Phi) - \tilde{f}^{ai}_{3}(\Phi)\star\left(\nabla^{\sigma}\Phi\star\tilde{f}^{aj}_{2}(\Phi)\right) \right]\overline{\epsilon}^{i}\epsilon^{j}.
\end{eqnarray}

The stability problem consists in finding the noncommutative functions $\tilde{f}^{ij}(\Phi)$, $\tilde{f}^{ai}_{2}(\Phi)$ and $\tilde{f}^{ai}_{3}(\Phi)$ for a given $\tilde{V}(\Phi)$ that ensure the positive-definiteness of Eq. (\ref{eq:ncdiv}).


\section{Stability conditions}\label{sec:stability}

The method to obtain the stability conditions is to identify Eq. (\ref{eq:ncdiv}) with Eq. (\ref{eq:gwi}). So to ensure that Eq. (\ref{eq:ncdiv}) is positive, its last five terms must vanish. Notice that this system of equation is very difficult to solve, but it can be simplified if one assumes that the indices $i,j,a$ are single-valued. 

The task is to examine each term at the R. H. S. of Eq. (\ref{eq:ncdiv}). The first term is positive definite since the matter fields satisfy the dominant energy condition. Choosing the time direction as the direction orthogonal to $\Sigma$ and using Eq. (\ref{eq:witten}), one obtains that the second and the third terms are $-4g^{mn}\left(\hat{\nabla}_{m}\epsilon\right)^{\dagger}\star\hat{\nabla}_{n}\epsilon + \left(\delta\chi\right)^{\dagger}\star\delta\chi$. By the assumption that $\theta^{\mu\nu}$ is covariantly constant, this sum is positive definite if one chooses the conditions:

\beq
\theta^{\mu\nu}\nabla_{\nu}\hat{\nabla}_{n}\epsilon =0, \;\;\theta^{\mu\nu}\nabla_{\nu}\delta\chi =0. \label{eq:condchi}
\eeq

Any noncommutative function $\tilde{h}(\Phi)$ can be expanded up to second order in $\theta$ as $\tilde{h}(\Phi)=h+i\theta^{\mu\nu}h_{\mu\nu}+\theta^{\alpha_{1}\beta_{1}}\theta^{\alpha_{2}\beta_{2}}h_{\alpha_{1}\alpha_{2}\beta_{1}\beta_{2}}$, where $h$ is a function of $\Phi$, $h_{\mu\nu}$ is an antisymmetric function of $\Phi$ and its derivatives, and so on. To examine each term in Eq. (\ref{eq:ncdiv}), one has to expand the functions of $\Phi$ up to second order, and the coefficients of each term in the expansion must vanish. This derivation has been fully carried on in Ref. \cite{Bertolami:2008zv}. As an example, one works in detail the term proportional to $\overline{\epsilon}\Gamma^{\sigma\alpha\beta}\epsilon$, which is absent in the commutative case discussed in Ref. \cite{Bertolami:1987wj}. It gives, after using that $\Gamma^{\sigma\alpha\beta}$ is totally antisymmetric and a formula found in  the Appendix of Ref. \cite{Bertolami:2008zv},

\beq
\frac{i\theta^{\mu\nu}}{2}f_{2}^{2}\nabla_{\mu}\nabla_{\alpha}\Phi\nabla_{\nu}\nabla_{\beta}\Phi\overline{\epsilon}\Gamma^{\sigma\alpha\beta}\epsilon-\theta^{\alpha_{1}\beta_{1}}\theta^{\alpha_{2}\beta_{2}}f_{2}f_{2\;\alpha_{2}\beta_{2}}\nabla_{\alpha_{1}}\nabla_{\alpha}\Phi\nabla_{\beta_{1}}\nabla_{\beta}\Phi\overline{\epsilon}\Gamma^{\sigma\alpha\beta}\epsilon,
\eeq

\noindent which vanishes if one chooses the condition

\beq\label{eq:condderphi}
\theta^{\mu\nu}\nabla_{\mu}\nabla_{\alpha}\Phi =0.
\eeq

The term proportional to $\overline{\epsilon}\epsilon$ vanishes if one uses condition Eq. (\ref{eq:condderphi}). From Eq. (\ref{eq:condderphi}) and formulas of the Appendix of Ref. \cite{Bertolami:2008zv}, the term proportional to $\overline{\epsilon}\gamma^{\alpha}\epsilon$ reads, as the coefficients of every order in the noncommutative parameter must vanish,

\beq\label{eq:f2}
f_{2}=\sqrt{\kappa}\;\;\;\;\;\;f_{2\;\mu\nu}=0\;\;\;\;\;\;f_{2\;\alpha_{1}\alpha_{2}\beta_{1}\beta_{2}}=0,
\eeq

\noindent and thus that $\tilde{f}_{2}(\Phi)=\sqrt{\kappa}$. For the term proportional to $\overline{\epsilon}\gamma^{\sigma}\epsilon$ one must assume that $\tilde{f}(\Phi)=a+b\Phi\star\Phi$, where constants $a$ and $b$ are obtained by the boundary conditions of the system of equations \cite{Bertolami:1987wj}. Using Eq. (\ref{eq:condderphi}) one finds the following equations

\beq
-f_{3}^{2}+2\kappa V(\Phi) + 6\kappa^{2}f^{2} =0, \;\;f_{3}f_{3\;\mu\nu}=0, \;\; f_{3\;\alpha_{1}\alpha_{2}\beta_{1}\beta_{2}}=\frac{f_{3\;\alpha_{1}\beta_{1}}f_{3\;\alpha_{2}\beta_{2}}}{2f_{3}}. \label{eq:f3a}
\eeq 

\noindent This system of equations yield

\beq\label{eq:f3munu}
f_{3\;\mu\nu}=0,\;\;\;f_{3\;\alpha_{1}\alpha_{2}\beta_{1}\beta_{2}}=0.
\eeq

Finally, the term proportional to $\overline{\epsilon}\Gamma^{\sigma\alpha}\epsilon$ gives

\beq
4\kappa\left(\frac{df}{d\Phi}\right)-2f_{2}f_{3} =0. \label{eq:derphif} 
\eeq

The stability conditions are obtained after solving the system of equations

\beq
2\sqrt{\kappa}\left(\frac{df}{d\Phi}\right)=f_{3}, \;\; -f_{3}^{2}+2\kappa V(\Phi) + 6\kappa^{2}f^{2} =0. \label{eq:stab2}
\eeq

This is precisely the set of equations for the commutative case for a quartic potential solved in  Ref. \cite{Bertolami:1987wj}. One concludes that the stability conditions for a scalar with a noncommutative potential are not affected by noncommutativity.

After obtaining the stability conditions one can verify the consistency of Eqs. (\ref{eq:condchi}). One shows that $\theta^{\mu\nu}\nabla_{\nu}\hat{\nabla}_{n}\epsilon =0 $ and $\theta^{\mu\nu}\nabla_{\nu}\delta\chi = 0$, using the fact that noncommutativity does not act on spinors and Eq. (\ref{eq:condderphi}).


\section{Space-time with horizons}\label{sec:bhspt}

To investigate black hole type spaces, the divergence theorem must be changed to include horizons, that is

\beq\label{eq:sggt}
\frac{1}{2}\oint_{S} \hat{E}^{\sigma\alpha}dS_{\sigma\alpha} - \frac{1}{2}\oint_{H} \hat{E}^{\sigma\alpha}dS_{\sigma\alpha} = \int_{\Sigma}\nabla_{\alpha}\hat{E}^{\sigma\alpha}d\Sigma_{\sigma},
\eeq

\noindent where $H$ is a two-surface that denotes the horizon. An orthonormal tetrad frame $\{e_{\hat{\mu}}\}$ is introduced at the horizon \cite{Straumann:1984xf}, such that: $e_{\hat{0}}$ is normal to $\Sigma$, $e_{\hat{1}}$ is normal to $H$ and $e_{\hat{A}}$ ($A=2,3$) are tangent to $H$. Using this coordinates one has to evaluate

\beq
\oint_{H} \hat{E}^{\hat{0}\hat{1}}dS_{\hat{0}\hat{1}}.
\eeq
                              
One omits for brevity the hat on the indices. One gets\footnote{The full derivation is presented in Ref. \cite{Bertolami:2008zv}.} after using the Witten's condition and the spinor identities: $\nabla_{b}\epsilon= ^{(3)}\!\!\nabla_{b}\epsilon+\frac{1}{2}K_{ab}\gamma^{0}\gamma^{a}\epsilon$ and ${ }^{(3)}\!\nabla_{A}\epsilon={ }^{(2)}\!\nabla_{A}\epsilon - \frac{1}{2}J_{AB}\gamma^{1}\gamma^{B}\epsilon$, where $^{(3)}\!\!\nabla_{b}$ and ${ }^{(2)}\!\nabla_{A}$ are respectively the intrinsic three-dimensional covariant derivative on $\Sigma$ and the intrinsic two-dimensional covariant derivative on $H$; $K_{ab}$ is the second fundamental form on $\Sigma$ and $J_{AB}$ is the second fundamental form on $H$.  Thus,   

\beq\label{eq:e01h2}
\left.\hat{E}^{01}\right|_{H}=\epsilon^{\dagger}\left[2\gamma^{1}\gamma^{A}\mathcal{D}_{A} - \left(J + \left(K + K_{11}\right)\gamma^{1}\gamma^{0}\right) + 2i\kappa \tilde{f}(\Phi)\gamma^{1} \right]\epsilon + \mbox{h. c.}\;\;\;\;,
\eeq 

\noindent where $\mathcal{D}_{A} \equiv \left({ }^{(2)}\!\nabla_{A} - \frac{1}{2}K_{1A}\gamma^{1}\gamma^{0} \right)$, $K=K^{a}_{\;\;\; a}$ and $J=J^{A}_{\;\;\;A}$. To further proceed, one must introduce another restriction over the spinor fields on $H$ \cite{Gibbons:1982jg}: $\gamma^{1}\gamma^{0}\epsilon =\epsilon$. Eq. (\ref{eq:e01h2}) can be written as

\beq\label{eq:e01h3}
\left.\hat{E}^{01}\right|_{H}=\epsilon^{\dagger}\left[2\gamma^{1}\gamma^{A}\mathcal{D}_{A} - \left(J + K + K_{11}\right)\right]\epsilon + 2i\kappa \tilde{f}(\Phi)\epsilon^{\dagger}\gamma^{1}\epsilon + \mbox{h. c.} \;\;\;\; .
\eeq 

Notice that $\left(J + K + K_{11}\right)=-\sqrt{2}\psi$, where $\psi$ is the expansion scalar \cite{Straumann:1984xf}, which vanishes if $H$ is an apparent horizon. Using the condition $\gamma^{1}\gamma^{0}\epsilon =\epsilon$, one sees that $\gamma^{1}\gamma^{0}$ anticommutes with $\gamma^{1}\gamma^{A}\mathcal{D}_{A}$ and with $\gamma^{1}$, so $\epsilon^{\dagger}\gamma^{1}\gamma^{A}\mathcal{D}_{A}\epsilon=0$ and $\epsilon^{\dagger}\gamma^{1}\epsilon=0$. If one chooses the boundary $H$ to be an apparent horizon, one finds

\beq
\oint_{H} \hat{E}^{\hat{0}\hat{1}}dS_{\hat{0}\hat{1}}=0,
\eeq

\noindent which shows that if one considers black hole type spaces, thus this does not alter the stability conditions obtained in the previous section.

\section{Conclusions}\label{sec:conclusions}

In this contribution the stability of noncommutative scalar fields coupled to gravity has been investigated. The proposed model consists in introducing the noncommutativity via a Moyal product adapted to curved spaces and an associativity condition $\theta^{\mu\nu}\nabla_{\nu}\Phi=0$. One finds that the stability conditions for a noncommutative polynomial scalar potential are the very ones obtained for the commutative case, already examined in Ref. \cite{Bertolami:1987wj}. It is shown that if $H$ is an apparent horizon then $\oint_{H}\hat{E}^{\sigma\alpha}dS_{\sigma\alpha}=0$, which proves that the stability conditions are not modified by the presence of horizons.

Finally, one stresses that the obtained stability conditions although similar to the commutative case are not by any means, trivial, as one might think at first sight. Indeed, one should notice that for the noncommutative case, the term proportional to $\overline{\epsilon}\Gamma^{\sigma\alpha\beta}\epsilon$, which is absent in the commutative case, gives origin to the new Eq.  (\ref{eq:condderphi}).


\begin{theacknowledgments}
The work of C. A. D. Z. is fully supported by the FCT (Funda\c{c}\~{a}o para a Ci\^{e}ncia e a Tecnologia, Portugal)
 fellowship SFRH/BD/29446/2006.
\end{theacknowledgments}

\end{document}